\documentclass[a4paper,12pt]{article}

\newcommand{\sect}[1]{\setcounter{equation}{0}\section{#1}}

\begin{document}
\topmargin 0pt \oddsidemargin 0mm

\renewcommand{\thefootnote}{\fnsymbol{footnote}}
\begin{titlepage}
\begin{flushright}
INJE-TP-02-06\\
hep-th/0210300
\end{flushright}

\vspace{5mm}
\begin{center}
{\Large \bf Holography and Entropy Bounds in Gauss-Bonnet Gravity}
 \vspace{12mm}

{\large Rong-Gen Cai\footnote{e-mail address: cairg@itp.ac.cn}$^1$
 and Yun Soo Myung\footnote{e-mail
 address: ysmyung@physics.inje.ac.kr}$^2$}
 \\
\vspace{10mm} {\em $^1$ Institute of Theoretical Physics, Chinese
Academy of Sciences, P.O. Box 2735, Beijing 100080, China \\
$^2$ Relativity Research Center and School of Computer Aided
Science, Inje University, Gimhae 621-749, Korea}
\end{center}

\vspace{20mm}
 \centerline{{\bf{Abstract}}}
 \vspace{5mm}
We discuss the holography and entropy bounds in Gauss-Bonnet gravity theory.
By applying a Geroch process to an arbitrary spherically symmetric black
hole, we show that the Bekenstein entropy bound always keeps its form  as
$S_{\rm B}=2\pi E R$, independent of gravity theories. As a result,
the Bekenstein-Verlinde bound also remains unchanged.  Along the Verlinde's
approach, we obtain  the Bekenstein-Hawking bound and Hubble bound, which
are different from those in Einstein gravity. Furthermore, we note that when
$HR=1$,  the three cosmological entropy bounds become identical as in the case of
Einstein gravity. But, the Friedmann equation in Gauss-Bonnet gravity
can no longer be cast to  the form of cosmological Cardy formula.

\end{titlepage}

\newpage
\renewcommand{\thefootnote}{\arabic{footnote}}
\setcounter{footnote}{0} \setcounter{page}{2}

\sect{Introduction}

According to the holographic principle~\cite{HP}, within a given volume $V$ the number
of degrees of freedom is bounded by a quantity proportional to the  surface area $A$
of the volume. This is obtained from the idea that the maximal entropy inside the
volume is given by the largest black hole that just fits inside the volume, while the
entropy of the latter obeys the Bekenstein-Hawking entropy formula $A/4G$, where $G$ is
the Newton constant. Thus, the holographic principle gives an entropy bound on matter
inside  the volume
\begin{equation}
\label{1eq1}
S \le \frac{A}{4G},
\end{equation}
which is called the holographic bound.

Fischler and Susskind~\cite{FS} were the first to consider entropy bound in the cosmological
setting. In a closed universe, the holographic bound in its naive form (\ref{1eq1}) is not
applicable because there is no boundary in the closed universe. On the other hand, the
argument leading to (\ref{1eq1}) assumes that it is possible to form a black hole filling the
whole volume. This is no longer valid in  the universe since the expansion rate $H$ of the universe
and the total energy in the universe restrict the maximal size of black hole~\cite{Verl}. Following Fischler and Susskind, it was argued that the maximal entropy inside the
universe is produced by black holes with size of Hubble horizon~\cite{Hubb}. The usual
holographic arguments  lead to the result  that the total entropy should be less than or
equal to the
Bekenstein-Hawking entropy of a Hubble horizon-sized black hole times the number of Hubble
regions in the universe. That is,  one has $S \le \beta HV/G$, where $V$ represents the volume of
the universe and $\beta$ is a pure coefficient.
This coefficient is fixed by Verlinde \cite{Verl} by using a local version of holographic
bound~\cite{FS,Bous}. This bound is called the Hubble entropy bound, which has the form
\begin{equation}
\label{1eq2}
S_{\rm H}= (n-1) \frac{HV}{4G},
\end{equation}
where $n$ stands for spatial dimensions of the universe. The Hubble bound is valid
 for a strongly self-gravitating universe ($HR\ge 1$). Except for the Hubble bound,
 Verlinde introduced other two entropy bounds~\cite{Verl}:
\begin{eqnarray}
\label{1eq3}
 {\rm Bekenstein-Verlinde\ bound}:&& S_{\rm BV}=\frac{2\pi}{n}ER
   \nonumber \\
 {\rm Bekenstein-Hawking\ bound}:&& S_{\rm BH}=(n-1)\frac{V}{4G_{n+1}R}.
 \end{eqnarray}
Here $E$ is the total energy of the matter filling the universe and $R$ is the scale factor.
The Bekenstein-Verlinde bound $S_{\rm BV}$ is the counterpart of the Bekenstein entropy
bound~\cite{Beke} in the cosmological setting~\cite{CMO}, which is believed to hold
for a weakly
self-gravitating universe ($HR \le 1$).  The Bekenstein-Hawking entropy bound
does not serve as an entropy bound, but acts as a criterion whether the universe is in a
weakly self-gravitating phase ($HR \le 1$) or in a strongly self-gravitating phase
($HR \ge 1$)~\cite{Verl}.  The Friedmann
equation of a ($n+1$)-dimensional, closed Friedmann-Robertson-Walker (FRW) universe is
\begin{equation}
\label{1eq4}
H^2 = \frac{16\pi G}{n(n-1)} \frac{E}{V} -\frac{1}{R^2},
\end{equation}
from which one can see that $S_{\rm BV} \le S_{\rm BH}$ for $HR \le 1$,
while $S_{\rm BV} \ge S_{\rm BH}$ for $HR \ge 1$. Clearly one has $S_{\rm BV}=S_{\rm BH}
=S_{\rm H}$ at the critical point $HR=1$. Furthermore, Verlinde found that with the three
cosmological entropy bounds, the Friedmann equation (\ref{1eq4}) can be cast to
\begin{equation}
\label{1eq5}
S_{\rm H}=\sqrt{S_{\rm BH} (2 S_{\rm BV}-S_{\rm BH})},
\end{equation}
the cosmological Cardy formula. This formula (\ref{1eq5}) has a close relation to the
Cardy-Verlinde formula describing the entropy of conformal field theories. For
more discussions, see \cite{Verl}.

We note that those discussions on the entropy bounds crucially depend on the
area entropy formula of black holes (\ref{1eq1}). However, it is well-known
that the area entropy formula of black holes holds only in Einstein gravity.
If some higher derivative curvature terms appear, for example, one has to include some
additional terms to the area entropy formula of black holes~\cite{TM}.
Therefore it would be interesting to see how those entropy bounds get
modified in higher derivative gravity theories. In this note we will discuss entropy
bounds in the  Gauss-Bonnet gravity, which belongs to a special class of higher
derivative gravity theories in the sense that the equation of  motion for the
Gauss-Bonnet gravity  contains no more than second derivatives of metric.

\sect{Bekenstein bound and Bekenstein-Verlinde bound}

Bekenstein was the first to consider the issue of maximal entropy for a
macroscopic system. He argued that for a closed system with total energy $E$,
which fits in a sphere with radius $R$ in three spatial dimensions, there exists an
upper bound on the entropy of the system
\begin{equation}
\label{2eq1}
S \le S_{\rm B} =2 \pi ER,
\end{equation}
which is called the Bekenstein entropy bound.  This bound is believed to be
valid for a system with limited self-gravity, which means that the gravitational
self-energy is negligibly small compared to its total energy. However, it is
interesting to note that this bound gets saturated for a ($3+1$) dimensional
Schwarzschild black hole, which of course is a strongly self-gravitating
object. Furthermore it was found that the form (\ref{2eq1}) is independent of
spatial dimensionality. That is, the form (\ref{2eq1}) keeps unchanged for any
dimensional object. This is obtained by considering a Geroch process in an arbitrary
dimensional Schwarzschild black hole and the generalized second law of black hole
thermodynamics~\cite{Bousso}. It is easy to show that for a higher ($n+1>4$) dimensional
Schwarzschild black hole, the Bekenstein entropy bound still holds, but it is not
saturated.

In deriving the Bekenstein entropy bound~\cite{Beke,Bousso}, black hole thermodynamics is used.
And  the thermodynamics of black holes  is dependent of gravity
theories under consideration.  In fact, we show here that the Bekenstein entropy
bound is independent of gravity theories. As a result, the Bekenstein-Verlinde bound has also
the same feature of independence of gravity theories.  Consider an
arbitrary, ($n+1$)-dimensional spherically symmetric black hole solution
\begin{equation}
\label{2eq2}
ds^2 =- e^{2\delta (r)}\left (1-\frac{2m(r)}{r^{n-2}}\right) dt^2
       +\left( 1-\frac{2 m(r)}{r^{n-2}}\right)^{-1}dr^2 +r^2d\Omega_{n-1}^2,
\end{equation}
where $\delta $ and $m$ are two continuous functions of $r$, It is assumed that
$e^{2\delta (r)} \ne 0$ in the whole spacetime. The black hole
horizon $r_+$ is determined by equation $1 -2m(r_+)/r^{n-2}_+=0$. The Hawking
temperature $T$ associated with the horizon is
\begin{equation}
\label{2eq3}
T=\frac{e^{\delta(r_+)}}{4\pi}\left(\frac{n-2}{r_+}-
  \frac{2m'(r_+)}{r_+^{n-2}}\right),
\end{equation}
where a prime denotes derivative with respect to $r$.  We denote by $M$  the mass of
 the black hole. According to the first law of black hole thermodynamics, which always
 holds because a black hole behaves as a thermodynamic system, one
has the entropy variation $\triangle S$ when the mass gets increase by a small
 amount $\triangle M$,
\begin{equation}
\label{2eq4}
\triangle S = T^{-1}\triangle M.
\end{equation}
Let us consider a Geroch process in the black hole background (\ref{2eq2}).
Suppose that one has a thermodynamic system with energy $E$ and $R$ being the radius
of the smallest ($n-1$)-sphere circumscribing the system. Now move this system from
infinity to a place just outside the horizon of the black hole (\ref{2eq2}), and drop
the matter into the black hole.

The mass added to the black hole is given by the energy $E$ of the system, which gets
redshifted according to the position of the center of mass at the drop-off point, at
which the circumscribing sphere almost touches the horizon. The center of mass can be
brought to within a proper distance $R$ from the horizon, while all parts of the system
still remain outside the horizon. Thus one needs to calculate the redshift factor at a proper
distance $R$ from the horizon~\cite{Bousso}.

Let $x$ be the radial coordinate distance from the horizon $x=r-r_+$. The redshift
factor near the horizon is given by
\begin{equation}
\label{2eq5}
\chi ^2(x)= e^{2\delta(r_+)}\left( \frac{n-2}{r_+}-\frac{2m'(r_+)}{r_+^{n-2}}
  \right) x
\end{equation}
up to the leading order of $x$. Near the horizon, the proper distance $R$ has a
relation to the coordinate distance $x$,
\begin{equation}
R=2 \sqrt{\frac{x}{(n-2)/r_+ -2m'(r_+)/r_+^{n-2}}}.
\end{equation}
Hence the absorbed mass is $\triangle M = E\chi(x)$. Substituting this
into (\ref{2eq4}), we find that the increased entropy of the black hole
is
\begin{equation}
\label{2eq7}
\triangle S = T^{-1} E \chi (x) = 2\pi ER.
\end{equation}
According to the generalized second law of black hole thermodynamics~\cite{Beke2},
which says that the total entropy of black hole and matter outside the black hole
never decreases in any physical process, we can immediately obtain the maximal
entropy of the system,
\begin{equation}
\label{2eq8}
S_m \le 2\pi ER.
\end{equation}
This is just the Bekenstein entropy bound (\ref{2eq1}).  From the above one can see
that we have neither specified  what the black hole solution (\ref{2eq2}) is,
nor in which gravity theory it is. Hence the resulting conclusion (\ref{2eq8})
is independent of gravity theories. This is an expected result since as stated
above, the Bekenstein
bound is valid only for systems with limited self-gravity, which implies that gravity
effect is negligible.   In addition, the Bekenstein-Verlinde bound $S_{\rm BV}$
in (\ref{1eq3}) is the counterpart of the Bekenstein bound in the cosmological setting.
Therefore we conclude that the Bekenstein-Verlinde bound is also independent of
gravity theories.

\sect{Hubble bound and Bekenstein-Hawking bound in Gauss-Bonnet gravity}

Now we consider the so-called Gauss-Bonnet gravity theory by adding the Gauss-Bonnet
term to the Einstein-Hilbert action,
\begin{equation}
\label{3eq1}
{\cal S} =\frac{1}{16\pi G}\int d^{n+1}x\sqrt{-g} \left ({\cal R}
    +\alpha ({\cal R}_{\mu\nu\gamma\sigma}
{\cal R}^{\mu\nu\gamma\sigma} -4 {\cal R}_{\mu\nu}{\cal R}^{\mu\nu} +{\cal R}^2)\right ),
\end{equation}
where $\alpha $ is a  constant. Here we exclude the case of $n=3$ since in that case the
Gauss-Bonnet term is a topological term. The static spherically symmetric black hole solutions
in (\ref{3eq1}) have been found in \cite{Deser,Whee}. The entropy of the black holes
has the expression~\cite{Myers,Cai}
\begin{equation}
\label{3eq2}
S =\frac{A}{4G}\left ( 1 +\frac{n-1}{n-3}\frac{2\tilde \alpha }{R^2}\right),
\end{equation}
where $\tilde \alpha =(n-2)(n-3)\alpha $, $A$ represents the horizon area of the black hole
and $R$ the horizon radius. Following Verlinde \cite{Verl}, in this section
we  ``derive" the Hubble bound for a closed FRW universe in the Gauss-Bonnet theory.

In \cite{Verl} Verlinde used a version of holographic bound proposed by Fischler and
Susskind \cite{FS} and subsequently developed by Bousso \cite{Bous}, which gives a
restriction of entropy flow
$S$ through a contracting light sheet: the entropy flow $S$ is less than or equal to $A/4G$,
where $A$ is the area of the surface from which the light sheet originates.  The
infinitesimal version of the holographic bound plays a crucial role in the ``derivation"
by Verlinde.  According to the infinitesimal version, for every ($n-1)$-dimensional
surface at time $t +dt$ with area $A+dA$, one has $dS \le dA/4G$. Here $dS$ represents
the entropy flow through the infinitesimal light sheets originating  at the surface
at $t +dt$ and extending back to time $t$, and $dA$ denotes the increase in area
between $t$ and $t +dt$. Obviously the
holographic bound is based on the area entropy formula of black holes. In our case,
the black hole entropy is given by (\ref{3eq2}). The infinitesimal version  is then
 changed to
\begin{equation}
\label{3eq3}
dS \le \frac{1}{4G}d\left[ A \left( 1 +\frac{n-1}{n-3}\frac{2\tilde \alpha }{R^2}\right)
  \right].
\end{equation}
In a ($n+1$) dimensional closed FRW universe, for a surface which is fixed in comoving
coordinates,  the area $A$ changes as a result of the expansion of the universe by
 an amount
\begin{equation}
\label{3eq4}
dA =(n-1)HA dt,
\end{equation}
where the relation $A \sim R^{n-1}$ has been used.  Choose one of two past light sheets
that originate at the surface: the inward or the outward going. The entropy flow through
this light sheet between $t$ and $t +dt$ is given by the entropy density $s=S/V$ times the
infinitesimal volume $A dt$ swept out by the light sheet. That is, one has
\begin{equation}
\label{3eq5}
dS =\frac{S}{V} A dt.
\end{equation}
Applying $A \sim R^{n-1}$ to (\ref{3eq3}), and then substituting (\ref{3eq3}) and
(\ref{3eq4}) into (\ref{3eq5}), we obtain
\begin{equation}
\label{3eq6}
S \le S_{\rm H} =(n-1) \frac{HV}{4G}\left (1+\frac{2\tilde \alpha}{R^2}\right),
\end{equation}
which is the Hubble entropy bound in the Gauss-Bonnet gravity. This is the main result
of ours.  If a cosmological
constant is added to the action (\ref{3eq1}), the entropy of black holes still has
the expression (\ref{3eq2})~\cite{Cai}.
Hence the Hubble entropy bound still takes the form (\ref{3eq6}) even if
 a cosmological constant is present in the Gauss-Bonnet gravity.

On the other hand, the Friedmann equation for the Gauss-Bonnet
gravity (\ref{3eq1}) is
\begin{equation}
\label{3eq7}
 H^2 +\frac{1}{R^2} +\tilde \alpha \left (H^2 +\frac{1}{R^2}\right)^2
 = \frac{16\pi G}{n(n-1)}\frac{E}{V},
\end{equation}
from which we see that when $HR=1$, the Bekenstein-Verlinde bound $S_{\rm BV}=
2\pi E R/n$ equals  the Hubble bound $S_{\rm H}$ given by (\ref{3eq6}).
This is a good check for our ``derivation" of the Hubble bound.
Furthermore, from the Friedmann equation (\ref{3eq7}), we find that the
Bekenstein-Hawking bound has the form
\begin{equation}
\label{3eq8}
S_{\rm BH} =(n-1)\frac{V}{4GR}\left( 1 +\frac{2\tilde \alpha}{R^2}\right).
\end{equation}
As in the case of Einstein theory \cite{Verl}, this Bekenstein-Hawking bound
(\ref{3eq8}) was obtained by identifying $S_{\rm BH}=S_{\rm BV}$ via the Friedmann
equation (\ref{3eq7}) at the critical point $HR=1$.  Hence at the critical point
the property that three cosmological entropy bounds  become identical in Einstein theory
persists
in the Gauss-Bonnet gravity. Inspecting the Friedmann equation (\ref{3eq8}), however,
we find that it can no longer be rewritten in the form (\ref{1eq5}), which might be related
to that the black hole entropy in higher derivative theories cannot be cast to the
Cardy-Verlinde formula~\cite{Cai2}.


\sect{Conclusion}

In summary this paper has initiated the study of holography in gravity theories with
higher derivative curvature terms. As a concrete  model, we have considered the Gauss-Bonnet
theory. We have shown that as expected, the Bekenstein bound and the Bekenstein-Verlinde
 bound keep the same forms as in Einstein theory,  while the Hubble bound and
Bekenstein-Hawking bound get modified. Along the Verlinde's approach, we have obtained
 expressions of the Hubble bound and Bekenstein-Hawking bound.  When the universe undergoes a transition from a weakly self-gravitating phase ($HR \le 1$) to a strongly
self-gravitating phase ($HR\ge 1$), the three  cosmological entropy bounds get matched at the
critical point $HR=1$, as in the case of Einstein gravity. However, the Friedmann
equation of the Gauss-Bonnet gravity cannot be rewritten in the form of the cosmological
Cardy formula.

\section*{Acknowledgment}

We thank H.W. Lee for useful discussions.
The work of R.G.C. was supported in part by a grant from Chinese
Academy of Sciences and a grant from Ministry of Education, PRC.
Y.S.M. acknowledges partial support from the KOSEF, Project
Number: R02-2002-000-00028-0.
R.G.C. is grateful to Relativity Research Center and School of Computer Aided
Science, Inje University for warm hospitality during his visit.

\end{document}